\documentstyle[twocolumn,aps,prl,epsfig]{revtex}

\begin{document}
\draft
\preprint{HEP/123-qed}

\wideabs{
\title{\Large\bf Most singular vortex  structures
in fully developed turbulence}
\author{\normalsize{S. I. Vainshtein}\\
{\small\it Department of Astronomy and Astrophysics, University
of Chicago, Chicago, Illinois 60637}\\}

\date{\today}
\maketitle
\begin{abstract}
\normalsize{
Using high Reynolds number experimental data, we  search for most dissipative,
most intense
structures. These structures possess a scaling
predicted by  log-Poisson model for the dissipation field
$\varepsilon_r$. The probability distribution function for
the exponents $\alpha$, $\varepsilon_r\sim e^{\alpha a}$, has been 
constructed, and compared with Poisson distribution. 
These new experimental data suggest that the most intense structures have
co-dimension less than 2. The log-Poisson statistics is compared with
log-binomial which follows from the random $\beta$-model. 
}
\end{abstract}
}

\narrowtext

Self-similar properties of turbulence, suggested by Kolmogorov
\cite{k41}, have been intensively studied for a long time. The theory predicted
simple scaling for the longitudinal velocity increments,  
$\langle |u(x+r)-u(x)|^p\rangle 
\sim r^{\zeta_p}$, where ${\zeta_p}=p/3$. It became clear, however, that there
are  corrections to this scaling, ${\zeta_p}=p/3+\tau(p/3)$, due to 
intermittency.  A
theory, incorporating the intermittency, the refined similarity 
hypothesis  \cite{kolm62}, links the
statistic of these corrections with the statistic of the dissipation field
$\varepsilon_r$, the energy dissipation averaged over a ball of size $r$. 
Namely,  $\langle \varepsilon_r^p\rangle \sim
r^{\tau(p)}$. Many models have been proposed to explain intermittency. It was
originally suggested that the statistics of  $\varepsilon_r$ is log-normal 
\cite{kolm62}. More
recently, She and L\'ev\^eque \cite{SL} (hereafter SL, see also \cite{D}, 
\cite{SW}, and recent study \cite{multiplier}) have proposed 
 log-Poisson statistics for the dissipation field, the theory resulting in a 
remarkable agreement with the experimentally found $\zeta_p$ in \cite{Benzi},
\cite{Benzi1}. These experimental exponents are obtained in Extended
Self-Similarity approach, which is useful because of extended scaling range.
In the following, we will also use these experimental exponents to compare with
 SL theory with slightly modified parameters. 

Consider the Poisson distribution (see, e.g., \cite{Poisson}):
\begin{equation}
P(\alpha,\xi)=e^{-\xi} {\xi^\alpha}/{\alpha!},~~\alpha=0,1,2,...,~~ 
\varepsilon_r=e^{\alpha a+b},
\label{Poisson}
\end{equation}
and let $a<0$, and therefore $b=\ln{\max{\varepsilon_r}}$. 
Calculating the moments, $\langle\varepsilon_r^p\rangle$, 
we note first that, as  $\langle\varepsilon_r\rangle=1$, we get,  
$b=\xi(1-e^a)$. 
Second, by definition, $\langle\varepsilon_r^p\rangle \sim r^{\tau(p)}$, and
therefore, $\xi=C\ln{(\ell/r)}$, 
$\ell$ being external scale, and $C$ is a constant. Finally, denoting $\gamma=
C(1-e^a)$, we obtain,
\begin{equation}
\tau(p)=C[1-(1-\gamma/C)^p]-p\gamma.
\label{spectrum}
\end{equation}

SL is recovered from (\ref{spectrum}) if $C=2$,  $\gamma=2/3$, so that
$a=\ln{(2/3)}$, $C$ being the
co-dimension of most dissipative structures, and $\gamma$ is defined by the
dissipation rate, i.e., inverse time-scale, $1/t_r\sim r^{-2/3}$ \cite{SL}.

The meaning of $C$ becomes even more clear directly from (\ref{Poisson}): 
the most intense
fluctuations correspond to $\alpha=0$ (as $a<0$), so that the probability
$P(\alpha=0)=e^{-\xi}=(r/\ell)^C=
(r/\ell)^{(D-H_0)}$, $D$ is dimension of space ($=3$). Thus the Hausdorff 
dimension for most dissipative structures
in SL theory, $H_0=1$, i.e., the structures are filaments.
On the other hand, using  expressions for $b$, $\xi$ and
$\gamma$, we now  rewrite (\ref{Poisson}) as follows,
\begin{equation} 
\varepsilon_r =
e^{\alpha a}\max{\varepsilon_r}=e^{\alpha
a}\left(\frac{r}{\ell}\right)^{-\gamma}. 
\label{maxima}
\end{equation}
 Putting $\alpha=0$
in (\ref{maxima}), we can see that the most intense structures are expected to
scale $\sim r^{-\gamma}$. 
 This  scaling predicted by the log-Poisson statistic
is proved to be possible to verify experimentally.
\begin{figure}
\psfig{file=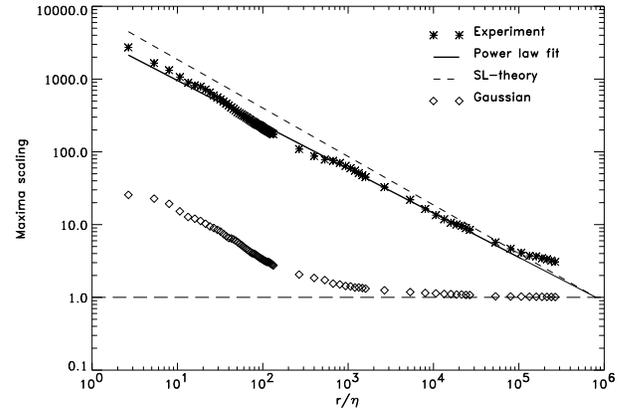,width=3.2in}
\caption{Scaling for most intense structures. The power law
fitting  of the experimental data (solid thick line) has been extended
to reach unity (solid line), where it is supposed to match with SL
scaling. The distances are
given in terms of Kolmogorov micro-scale $\eta$. }
\end{figure}

We used 10 million points of atmospheric data 
from Yale University, with an estimated Taylor microscale Reynolds 
number 9540 (courtesy of Sreenivasan).  The data are treated in  spirit of Taylor
hypothesis, that is, the time series is treated as one-dimensional cut of
the process.
 We denote $\omega=\partial_x u$, $u$ is longitudinal
velocity along the $x$-axis. Therefore, the dissipation, $\varepsilon \sim
\omega^2$, and we will deal with dimensionless dissipation $\varepsilon=
\omega^2/\langle \omega^2 \rangle$. In our case,
\begin{equation}
 \varepsilon_r=\frac{1}{r}\int_{x-r/2}^{x+r/2}\frac{\varepsilon(x')dx'}
{\langle \varepsilon \rangle},
\label{coarse}
\end{equation}
 and maxima of $\varepsilon_r$ can be measured. Figure 1
shows a remarkable scaling for this quantity for $4.5$ decades. The deviation
from SL is small, and we recall that SL suggest that there is no anomalous
scaling for $t_r$. This small deviation in Fig. 1 can be interpreted
\begin{figure}
\psfig{file=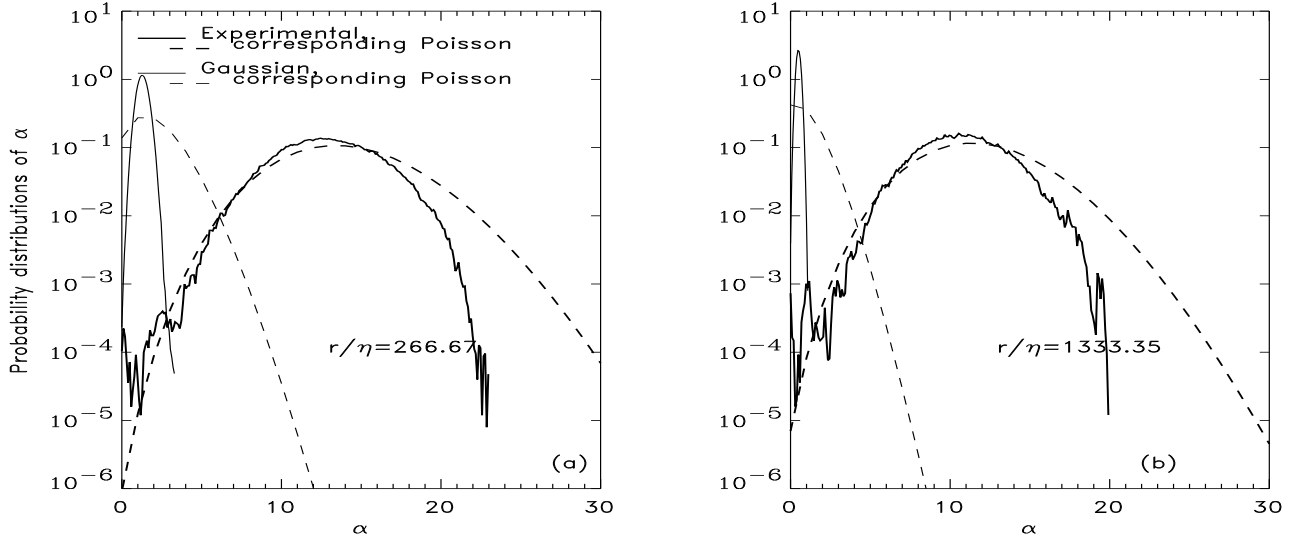,width=7in,height=3in}
\caption{$\rm Constructing~ PDF's ~for~ experimental~ and~ Gaussian
~ processes,~ and~ comparing~ them~
with~ corresponding~ Poisson~\\ processes.$
}
\end{figure} 
\noindent
as anomalous
persistence of the eddies, which is indeed observed \cite{persistence}, see also
discussion in \cite{our}.
The value of $\gamma$ is $0.61\pm 0.01$, only slightly smaller than $2/3$.
Note that maxima are not always predictable; e.g., for log-normal distribution,
the maximum could be anything. Another point: the measurements of maxima is
meaningful because the coarse-grain averaging (\ref{coarse}) is already present.
In order to compare with a ``regular" random process we generated a Gaussian
process $\omega_g$ with  correlation function coinciding with experimental,
i.e.,  $\langle \omega_g(x+r)\omega_g(x)\rangle=\langle
\omega(x+r)\omega(x)\rangle$. Then, the ``dissipation"
$\varepsilon^{(g)}=\omega_g^2$, and $\varepsilon^{(g)}_r=1/r
\int_{x-r/2}^{x+r/2}\varepsilon^{(g)}(x')dx'$. Corresponding calculation for 
the maxima are
reported in Fig. 1. If any scaling can be extracted from the Gaussian process, it
would be at large asymptotic distances, and the scaling is trivial, $\gamma=0$,
meaning no singularity.

Another testing of the theory is presented by direct measurements of the
exponents $\alpha$ in (\ref{maxima}). We may  construct a PDF for 
$\alpha$-distribution,  measuring
$\ln{\varepsilon_r}$, and taking $a=\ln{\{2/3\}}$, say. In fact, the
calculated value of $a$ is only slightly different from $\ln{\{2/3\}}$, see below,
and, as a result, the PDF plots with "true" $a$ are actually indistinguishable
from these with $a=\ln{\{2/3\}}$. In order to compare the experimental PDF with the
Poisson distribution we find $\xi$ from (\ref{maxima}),  $\xi=
\ln{\{\max{\varepsilon_r}\}}/(1-e^a)$, and hence the distribution 
(\ref{Poisson}) is unambiguously defined. Thus, instead of making use of
 inverse
Legendre transform of the spectrum (\ref{spectrum}), we provide direct
measurements of the exponents $\alpha$.

Figure 2 shows two typical PDF's for two distances. It is clearly seen that 
the SL theory is
well confirmed: the experimental PDF's are quite close to the corresponding
Poisson distributions. The deviation is observed only at large $\alpha$, i.e.,
(exponentially) small $\varepsilon_r$, for which the theory does not claim to
account for. The latter

{\vspace{3.5in}}
\noindent
 is constructed to account for asymptotically high 
moments, i.e., high values of $\varepsilon_r$. As to the Gaussian process, it is
clear that first, it is dramatically different from the experimental, and
second, it is quite different from corresponding Poisson distribution
(which is defined by $\xi_g=
\ln{\{\max{\varepsilon^{(g)}_r}\}}/(1-e^a)$). The
latter predicts much higher level of strong fluctuations (corresponding to low
$\alpha$) than actual Gaussian -- which is hardly surprising.

Returning to the Legendre transform, we note that, for comparison, we have to
express $\alpha!$ in (\ref{Poisson}) through Stirling formula, to get,
\begin{equation}
P=\frac{1}{\sqrt{2\pi\alpha}}e^{\alpha-\xi+\alpha\ln{\{\xi/\alpha\}}}=
\frac{1}{\sqrt{2\pi\alpha}}\left(\frac{r}{\ell}\right)^{D-H},
\label{Legendre}
\end{equation} 
where $D-H=C(1-y+y\ln{y})$, $y=(h+\gamma)/(C|\ln{\{1-\gamma/C\}}|)$, 
and $h=-\gamma-\alpha
|\ln{\{1-\gamma/C\}}|/\ln{\{r/\ell\}}$. In spite of the fact that Stirling
formula is  valid only for $\alpha\gg 1$, formula (\ref{Legendre})
for our parameters (in particular,
$\xi=14\div 16$) is quite close to (\ref{Poisson}) except for $\alpha \le 1$
(and, according to (\ref{Legendre}), $P\to\infty$  when $\alpha\to 0$).
Moreover, the exponent $D-H$ exactly coincides with that obtained from Legendre
transform, i.e., $D-H=D-\inf_p{\{ph+D-\tau(p)\}}$. Now, $\min{h}=-\gamma$,
corresponds to maximal excitation, according to (\ref{maxima}), while the
corresponding Hausdorff co-dimension $D-H(\min{h})=C$ \cite{SL}.

Although the experimental $\gamma$ is not that different from $2/3$, the value of
the other parameter $C$ is quite sensitive to that difference. 
In order to find $C$ we substitute  $\gamma$ from our measurements into
(\ref{spectrum}), and use computer routines to find a best fit for these data 
with free parameter $C$ and the exponents $\zeta_p^{(ESS)}$ from experiment 
\cite{Benzi}, \cite{Benzi1}.
 As a result, we find
$C=1.67$ and $a=-0.45$ (cf. $\ln{\{2/3\}}=-0.41$). With these parameters, the
 deviation of these calculated exponents $\zeta_p^{(e)}$ from the experimental 
exponents, $\sqrt{\langle(\zeta_p^{(e)}-\zeta_p^{(ESS)})^2\rangle}=0.0063$. 
To compare: for SL, $\sqrt{\langle(\zeta_p^{(SL)}-\zeta_p^{(ESS)})^2\rangle}=
0.0078$. The  $\zeta_p^{(e)}$ exponents seem to be ``better" than 
$\zeta_p^{(SL)}$, but considering that the experimental exponents have  
errors of about $\pm 1\%$ \cite{Benzi1}, we conclude that these exponents are
similar.  
Note that if we substitute in
(\ref{spectrum}) the value of $\gamma$ from the experiment, and put $C=2$, then,
for the obtained exponents, $\zeta_p^{(C=2)}$, we have, 
$\sqrt{\langle(\zeta_p^{(C=2)}-\zeta_p^{(ESS)})^2\rangle}=
0.050$, much too high. Figure 3 shows $\tau(p/3)$ from experiment, and for
different theories. It can be seen that all the curves collapse into one,
corresponding to the experiment, except that one with $\gamma$ from 
our measurements, and $C=2$. This illustrates that the data are indeed sensitive
to the measured $\gamma$, that is to its (small) difference from $2/3$.
 The codimension $C=1.67$ corresponds to $H_0=1.33$. This value of $H_0>1$ seems
to be consistent with the distinction between persistent vortical filaments and
the dissipative structures associated with regions of strong strain
\cite{vortex}.
That means that the most dissipative structures consist not only of 
filaments, but in part of sheets, or filaments convoluted into complex
structures, covering more than $1$ dimension. 
\begin{figure}
\psfig{file=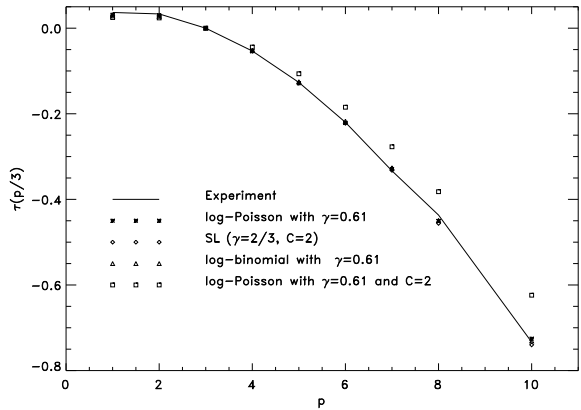,width=3.2in}
\caption{
Intermittency corrections from experiment  [\ref{Ben}],
 and from other 
theoretical models.}
\end{figure}

Note that, according to the
intersection theorem \cite{mandel}, dimension $H_0 \le 2$ cannot be detected in
1D measurements directly, and therefore our conclusion is inevitably indirect.
It would be important to measure the Hausdorff dimension in 3D simulations
directly. Another reason for that is the surrogacy issue \cite{shreeni}.

These statements about the dimensions of the most
intense structures can be also formulated for log-binomial distribution, and, it
is known that the Poisson process is a limit of the binomial distribution for
``rare events". In particular, the Poisson distribution can be obtained
from the random $\beta$-model \cite{random} by a suitable limiting process 
\cite{D}, \cite{B}. Let $\beta$ take two values, $W=\beta_1$ with probability
$x$, and $W=\beta_2$ with probability $1-x$,
and $\beta_1x+\beta_2(1-x)=1$ (in order to have $\tau(1)=0$).
 Let also $x\leq 1/2$, and $\beta_1\leq 1\leq
\beta_2$.  Then, on the $n$-th level, the distribution is binomial, that is,
$W_n=\varepsilon_n=\beta_1^m \beta_2^{n-m}$ with probability
 $( ^n_m)x^m (1-x)^{n-m}$. Hence, $\langle \varepsilon_n^p\rangle=
[x\beta_1^p+(1-x)\beta_2^p]^n$. Taking into account that
$n=\ln{(r/\ell)}/\ln{\Gamma}$,
$\Gamma$ being the ratio of successive scales, we obtain,
\begin{equation}
\tau(p)=
\frac{\ln{[x\beta_1^p+(1-x)\beta_2^p]}}{\ln{\Gamma}}.
\label{spectrum1}
\end{equation}
In \cite{D} and \cite{B}, $\Gamma$ was treated as a free parameter. It was shown
that, if $\Gamma=1-x/C$,  $\beta_1=1-\gamma/C$ and $x\to 0$, then
$\beta_2\approx 1+x\gamma/C$, and
(\ref{spectrum1}) reduces to (\ref{spectrum}). 

The most intense structures on $n$-th level,
\begin{equation}
\beta_2^n=\left(\frac{r}{\ell}\right)^{\ln{\beta_2}/\ln{\Gamma}}=
\left(\frac{r}{\ell}\right)^{-\gamma_\beta}, 
\label{maxima1}
\end{equation}
cf. 
(\ref{maxima}). On the other hand, the probability of these maxima, 
\begin{equation}
P=(1-x)^n=\left(\frac{r}{\ell}\right)^{\ln{(1-x)}/\ln{\Gamma}}=
\left(\frac{r}{\ell}\right)^{C_\beta}.
\label{probability}
\end{equation}
In particular, if $\Gamma=1-x/2$ and $x\to 0$, then $C_\beta=2$ \cite{B1}.

If we do not treat  $\Gamma$ as a
free parameter, we may consider $\Gamma=1/2^{1/D}$, say. As $1\leq \beta_2
\leq 2$, $\gamma_\beta$, according to (\ref{maxima1}), satisfies 
$0\leq \gamma_\beta \leq
D$. And, according to (\ref{probability}), $0\leq
C_\beta~(=\ln{(1-x)}/\ln{\Gamma})\leq D$. Thus, both $\gamma_\beta$ and 
$C_\beta$ satisfy
requirements for codimensions.
 A particular
case $x=1/2$ corresponds to the model proposed in \cite{MS}. Then, according to
(\ref{probability}), $P=(r/\ell)^D$, i.e., the Hausdorff dimension $H_0$ is
$=0$, while $\gamma_\beta$, defined from (\ref{maxima1}), 
$=D\ln{\beta_2}/\ln{2}$. 
The case $\beta_1=0$ returns us to the $\beta$-model \cite{Frisch}. In that case,
$\beta_2=1/(1-x)$, and $\gamma_\beta=C_\beta=D\ln{(1-x)}/\ln{(1/2)}$. 

The log-binomial 
distribution generally cannot be reduced to the log-Poisson PDF: in particular,
even if $x$ is small, and $\Gamma=1/2^{1/D}$, then, $\tau(p)\sim x\to 0$, and thus
the intermittency is negligible . Then, at first sight, the log-binomial
distribution is not suitable for our purpose. Indeed, according to the central
limit theorem, the binomial distribution is asymptotically normal at $n\gg 1$,
and the log-normal distribution has unsurmountable shortcomings \cite{book},
\cite{Fbook}.
Nevertheless, the spectrum 
(\ref{spectrum1}) does not even look like log-normal
(for which $\tau(p)=-(\mu/2)p(p-1)$)
and rather behaves like log-Poisson for $p\gg 1$. Indeed, according to 
(\ref{spectrum1}), for $p\gg 1$,
\begin{equation}
\tau(p)=C_\beta +C_1\beta^p-p\gamma_\beta,~~
C_1=\frac{x}{(1-x)\ln{\Gamma}}<0, 
\label{big}
\end{equation}
$\beta=\beta_1/\beta_2$.
This spectrum resembles (\ref{spectrum}); and the constants in (\ref{big})
happen to be numerically close to corresponding numbers in (\ref{spectrum}).
The reason for such a dramatic difference with log-normal distribution is as
follows. For binomial distribution,
\begin{equation}
\langle \varepsilon_n^p\rangle=\beta_2^{np}\sum_{m=0}^n  
(^n_m)x^m(1-x)^{n-m}\delta^m,
\label{sum}
\end{equation}
$\delta=\beta^p \ll 1$
for large $p$. Then $\delta^m$ decreases dramatically with
increasing $m$, and therefore the terms of the sum (\ref{sum}) of maximal probability, at
$m\sim xn$, where  normal distribution if formed, do not
contribute substantially. In contrast, only the first few terms of this sum
(responsible for "rare" and very intense events) really contribute. 
Thus, effectively, the distribution works like a Poisson
distribution. To see this explicitly, consider a probability distribution
$(^n_m)x^m(1-x)^{n-m}\delta_0^m/A$, where $A$ is a normalization constant,
$A=(x\delta_0 +1-x)^n$, and $\delta_0=(\beta_1/\beta_2)^{p_0},~~ p_0\gg 1$.
Then, for large $n$ we express the factorials entering the binomial 
coefficients through Stirling formula (except for $m!$, because $m$ is not
necessarily large), to get,
\begin{equation}
P_1(m)=\frac{1}{A}(^n_m)x^m(1-x)^{n-m}\delta_0^m\approx e^{-\xi_0}
\frac{\xi_0^m}{m!},
\label{Poisson1}
\end{equation}
 where $\xi_0=n\delta_0 x/(1-x)$. For $p\ge p_0$, the sum (\ref{sum}) can
be written as $A\beta_2^{np}\sum_{m=0}^\infty P_1(m)\delta^{m(p-p_0)}$, and thus the
distribution  effectively corresponds to the  Poisson distribution.
It is also important to note that log-normal distribution has an infinite 
maximum,
unlike the log-binomial. To see this, recall that $\tau(p)=-d_p(p-1)$,
$d_p=D-D_p$, where $D_p$ are so-called generalized dimensions \cite{generalized}.
Therefore, $\gamma=d_\infty=\lim_{p \to\infty}{\{-\tau(p)/(p-1)\}}$. For the
log-normal process, $d_p=(\mu/2)p$ and $\gamma\to \infty$.

If we take the random $\beta$ model ``for real", that is, consider the Poisson
distribution as an approximation to the binomial, as in (\ref{Poisson1}), then
we are dealing with (\ref{spectrum1}) with $\gamma_\beta$ given from our 
measurements
(so that $\beta_2$ is defined according to (\ref{maxima1})). And the second
parameter, the co-dimension $C_\beta$ from (\ref{probability}), can be found 
from the
best fit with the experimental data \cite{Benzi}, \cite{Benzi1} - 
using computer routines analogous to
those we used for the
Poisson distribution (see above). The resulting $C_\beta=1.58$, and we obtain
exponents $\zeta_p^{(bi)}$, for which
$\sqrt{\langle(\zeta_p^{(bi)}-\zeta_p^{(ESS)})^2\rangle}=0.0064$, quite
satisfactory.
 Indeed,
the corresponding $\tau(p/3)$  depicted in Fig. 3 is indistinguishable
from other approximations which collapse to the experimental data. 

In conclusion, one of the predictions of SL theory about the scaling of maxima
$\sim r^{-\gamma}$ is experimentally confirmed. This makes it possible to make a 
better estimate of
the intense structures geometry in fully developed turbulence. The PDF's of the
exponents of the dissipation field are compared with the log-Poisson
distribution to show a good agreement with the theory. The log-Poisson statistic
can be considered as a limiting case for the log-binomial distribution appearing
in random $\beta$-model. We estimated the parameters of the log-binomial
distribution  with $\gamma$ found in our measurements, and to fit the exponents
for the structure functions found elsewhere. We conclude that the estimated
Hausdorff co-dimension of the most intense structures is less than 2.
 
I thank K. R. Sreenivasan 
and B. Dhruva for  sharing with me the data of 
atmospheric turbulence. I appreciate numerous comments made by S. Boldyrev, Z.
Miki\'c, and R. Rosner.

\end{document}